\documentclass[english]{article}
\usepackage[T1]{fontenc}
\usepackage[latin9]{inputenc}
\usepackage{geometry}
\usepackage{amssymb}
\usepackage{amsmath}
\usepackage{graphicx}
\usepackage{setspace}
\usepackage{esint}
\usepackage{babel}
\usepackage{authblk}
\usepackage{hyperref}
\usepackage[dvipsnames]{color}

\begin{document}

\title{Current-phase relations of InAs nanowire Josephson junctions: from interacting to multi-mode
regimes}
\author[1,2*]{Sean Hart}
\author[1,2,3*]{Zheng Cui}
\author[4]{Gerbold M\'enard}
\author[4]{Mingtang Deng}
\author[5]{Andrey Antipov}
\author[5]{Roman M. Lutchyn}
\author[4,6]{Peter Krogstrup}
\author[4]{Charles M. Marcus}
\author[1,2,3]{Kathryn A. Moler}

\affil[1]{Stanford Institute for Materials and Energy Sciences, SLAC National Accelerator Laboratory, Menlo Park, California 94025, USA.}
\affil[2]{Geballe Laboratory for Advanced Materials, Stanford University, Stanford, California 94305, USA.}
\affil[3]{Department of Applied Physics, Stanford University, Stanford, California 94305,
USA.}
\affil[4]{Center for Quantum Devices and Microsoft Quantum Lab - Copenhagen, Niels Bohr Institute, University of
Copenhagen, Copenhagen DK-2100, Denmark.}
\affil[5]{Station Q, Microsoft Research, Santa Barbara, California 93106-6105, USA.}
\affil[6]{Microsoft Quantum Materials Lab, Copenhagen, Denmark.}
\affil[*]{These authors contributed equally to this work.}

\maketitle

\begin{abstract}

Gate-tunable  semiconductor-superconductor  nanowires  with  superconducting  leads  form  exotic  Josephson  junctions  that  are  a  highly  desirable  platform  for  two  types  of  qubits:  those  with  topological  superconductivity  (Majorana  qubits)  and  those  based  on  tunable  anharmonicity  (gatemon qubits).  Controlling  their  behavior,  however,  requires  understanding  their  electrostatic  environment  and  electronic  structure.  Here  we  study  gated  InAs  nanowires  with  epitaxial  aluminum  shells.  By  measuring  current-phase  relations  (CPR)  and  comparing  them with analytical and numerical  calculations,  we  show  that  we  can  tune  the  number  of  modes,  determine  the  transparency  of  each  mode,  and  tune  into  regimes  in  which  electron-electron  interactions  are  apparent, indicating the presence of a quantum dot. To take into account electrostatic and geometrical effects, we perform  microscopic  self-consistent  Schrodinger-Poisson  numerical  simulations, revealing the energy spectrum of Andreev states in the junction as well as their spatial  distribution.  Our work systematically demonstrates the effect of device  geometry,  gate  voltage  and   phase bias on mode behavior, providing new insights into ongoing experimental efforts  and  predictive  device  design.
\end{abstract}

\section*{Introduction}

Studies of hybrid structures of superconductors and semiconducting nanowires have recently spurred advances in fundamental physics, materials science, and technology \cite{lutchyn2018review}.  A growing body of experimental evidence supports proposals that nanowire-superconductor hybrids can host Majorana modes, the building blocks in schemes for topological quantum computing \cite{kitaev2001unpaired, lutchyn2010majorana, oreg2010helical, mourik2012signatures, deng2012anomalous, das2012zero, rokhinson2012fractional, churchill2013superconductor, finck2013anomalous, albrecht2016exponential, deng2016boundstate, zhang2018quantized}.  Gate-tunable nanowire Josephson junctions have also been integrated into a cavity-QED architecture, providing electrostatic control of the qubit transition frequency \cite{larsen2015nwtransmon, delange2015nwtransmon}.  Improving the design and operation of these devices will rely on understanding their electrostatic environment and electronic structure, with several recent works examining mode behavior in wire devices \cite{vanwoerkom2017uwave, spanton2017cpr, goffman2017mar, kringhoj2018anharmonicity, hays2018absdynamics,tosi2018andreev}.  Moreover, electron-electron interactions, often overlooked in both experiment and theory, should also be considered as they are expected to play an important role, particularly in one or fewer dimensions.

Here we report measurements of the current-phase relation (CPR) of a Josephson junction based on an InAs nanowire, carried out using a scanning Superconducting QUantum Interference Device (SQUID) microscope.  The amplitude and phase dependence of the CPR are sensitive to the transparency and number of modes contributing to supercurrent in the junction.  Microscopic regimes with one or multiple modes can be distinguished through comparison to a simple analytic model, by a model-independent analysis, and by comparison to numerical simulations.  The good agreement between our simulations and experiment provides a way to visualize the spatial arrangement of states in the wire and their response to electrostatic gating.  In the regime of low electron density in the junction, we find evidence for interactions, which modify both the spectrum and the phase-dependence of current flow in the device (i.e., lead to a sign reversal of the supercurrent). Our simulation of the device behavior, combined with our understanding and control of both interactions and anharmonicity, provides new insight into recent experiments and allows for more deterministic device design.

\section*{Measuring the current-phase relation of an InAs nanowire Josephson junction}

\begin{figure}
   \centering
   \includegraphics[width=\textwidth]{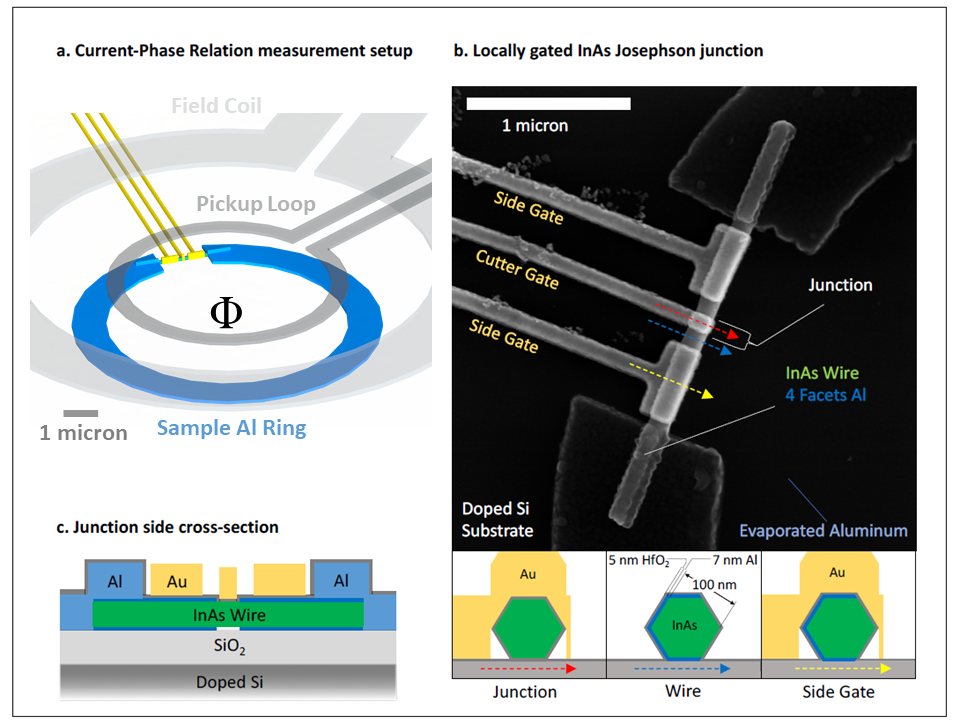}
   \caption{Experimental setup to measure the current-phase relation of an InAs nanowire-based Josephson junction.  a)  The sample ring consists of an evaporated aluminum ring (blue), which is interrupted by an InAs nanowire (green with yellow gates) to form a Josephson junction.  To measure the current-phase relation of the junction, the sensing head of a SQUID microscope is positioned above the sample ring.  A current bias applied to the field coil (light grey) tunes the magnetic flux through the sample ring, providing control of the phase difference across the junction.  This phase difference determines the supercurrent in the junction via its current-phase relation. The supercurrent circulates in the sample ring, leading to a signal which is measured by the pickup loop (dark grey).  Drawing is to scale.  b) A scanning electron micrograph depicts a top view of the portion of the sample ring spanned by the InAs nanowire.  As depicted in the bottom panels, the nanowire has a hexagonal cross-section, and is coated epitaxially on four facets with a 7 nm layer of aluminum.  This epitaxial aluminum layer is chemically removed along a 100 nm span beneath the cutter gate, while the wire is isolated from the gate with a 5 nm layer of HfO$_2$ (red cross section).  The blue and yellow cross-sections indicate the structure of the wire in ungated sections and in sections with a side gate. c) A cross-sectional side view of the wire shows the layer structure of the device.  The sample ring rests on a doped silicon substrate, providing the ability to gate the device globally from below.}
   \label{fig:one}
\end{figure}

To investigate the CPR of the InAs nanowire Josephson junction, we fabricated a superconducting ring consisting of a 100 nm thick annular film of evaporated aluminum bridged by a hybrid epitaxial aluminum--InAs nanowire (Figure 1a). The Josephson junction is located at the center of the nanowire. To measure the current-phase relation, we positioned the pickup loop and field coil of a SQUID microscope approximately 1 micron above the sample ring \cite{kirtley2016scanning}.  When a current is applied to the field coil, a magnetic flux $\Phi$ is generated in the sample ring.  This flux directly tunes the phase difference $\phi$ across the nanowire junction, leading to a supercurrent in the junction due to its current-phase relation.  The supercurrent circulates in the sample ring and is measured as a flux in the pickup loop \cite{jackel1974direct,sochnikov2013direct,sochnikov2015nonsinusoidal}.  By modeling the SQUID-sample geometry and calibrating the mutual inductance against the periodicity of the current-phase relation, we convert the flux signal into a Josephson supercurrent in the ring (See Supplemental Material \ref{sup:mutual}).

The hybrid nanowires are grown by Molecular Beam Epitaxy \cite{krogstrup2015epitaxy}. Au-catalyzed InAs nanowires are first grown via the vapor-liquid-solid method followed by a low temperature in-situ growth step of 7 nm aluminum on four of the six side facets (Figure \ref{fig:one}b). The four facet coverage is realized in two steps of two-facet growth with an intermediate 120 degree rotation of the nanowire orientation with respect to the atomic aluminum beam, which ensures a flat continuous aluminum film on all four facets.  This epitaxial aluminum makes electrical contact with the evaporated aluminum film to close the ring.  In the center of the nanowire, the epitaxial aluminum is chemically removed over a span of 100-150 nm to form the junction.  The junction lies under a Ti/Au ``cutter'' gate, while two additional side gates overlap the wire in regions still contacted with epitaxial aluminum (see Figure 1c).  A 5 nm layer of HfO$_2$ isolates the gates and the wire.  The side gates are expected to be mostly screened by aluminum, but can still affect the wire via fringe electric fields.  Finally, the sample ring sits on a doped silicon substrate capped with a thin layer of oxide, allowing the wire to be gated from below.  We found that the cutter gate and back gate have qualitatively similar effects, indicating that both gates primarily affect the same junction region (under the cutter gate) due to screening from the epitaxial aluminum (See Supplemental Material \ref{sup:gates}). Hence, we focus mainly on the modulation of the CPR by the cutter gate and side gates.

\section*{Evolution of the current-phase relation with the cutter gate voltage}

\begin{figure}
   \centering
   \includegraphics[width=\textwidth]{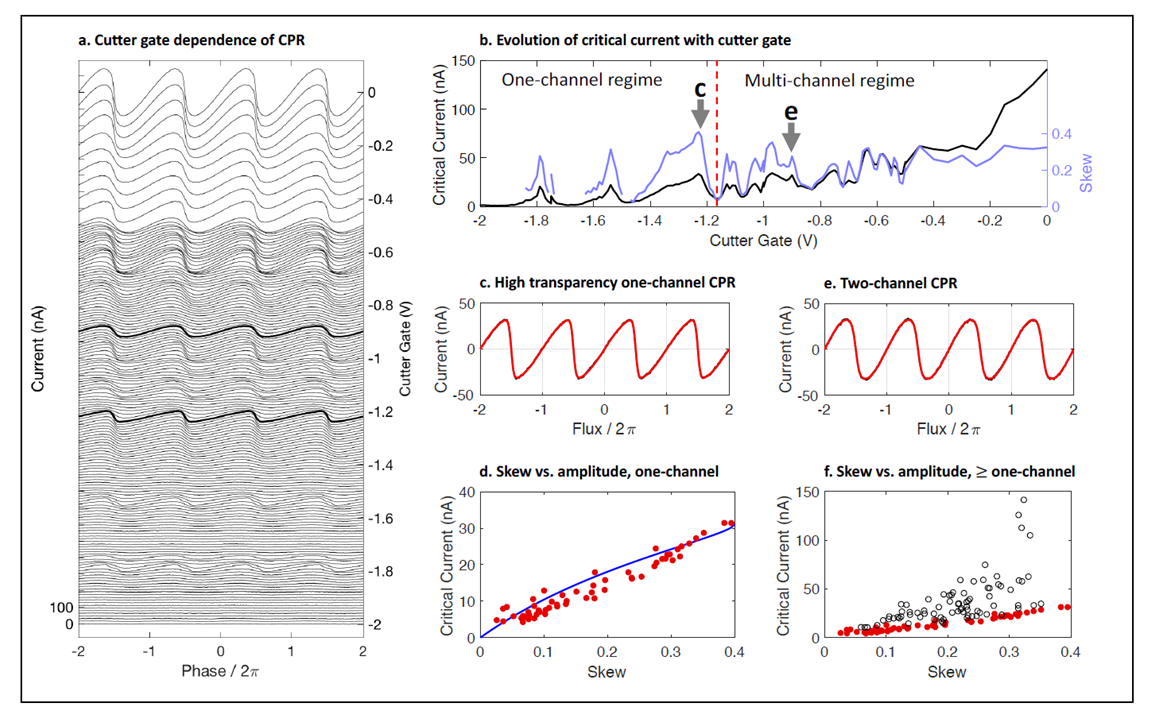}
   \caption{CPR measurements probe the evolution of supercurrent-carrying modes with gate voltage.  a) The amplitude and phase-dependence of the CPR evolve as the voltage applied to the cutter gate decreases from 0 to $-$2 volts, corresponding to changes in the structure of the Andreev bound states as a function of applied electric field.  The CPR data is displaced vertically to correspond with the gate voltage axis at right. b) The dependence of the CPR amplitude on cutter gate voltage reveals several distinct regimes of device behavior.  At the most negative gate voltages, a single channel contributes to the CPR and recurring peaks in amplitude are observed.  When the cutter gate voltage increases above approximately $-$1.16 volts (red dashed line), the CPR amplitude begins to fluctuate and multiple channels contribute to supercurrent.  Finally at the most positive cutter gate voltages, the CPR amplitude and channel number increase monotonously with gate voltage.  c) The single-channel regime can be identified by fitting data (black) to a simple model for the CPR (red).  d) In the single-channel regime, the CPR amplitude and skew are correlated.  e) The multi-channel regime can also be identified by fitting data (black) to the model (red). In this case the fit indicates two contributing channels.  The plotted data in (c) and (e) are highlighted by dark curves in (a) and arrows in (b).  f) In the multi-channel regime, the relationship between amplitude and skew is no longer correlated as with a single channel, but still constrains modeling.}
   \label{fig:two}
\end{figure}

To explore the effect of local gating on the nanowire junction, we measured the CPR at a series of voltages $V_C$ applied to the cutter gate (Figure \ref{fig:two}a).  As the cutter gate voltage is tuned from $-2$ to $0$ V, the amplitude of the CPR evolves through several qualitatively distinct regimes.  At the most negative gate voltages, the onset of supercurrent gives way to a set of recurring peaks in the critical current (plotted in black in Figure \ref{fig:two}b).  As the gate voltage increases above $V_C \approx -1.16$ V, the critical current begins to fluctuate with gate voltage.  Finally, above $V_C \approx -0.45$ V, the fluctuations diminish and the critical current begins to increase more steeply with gate voltage.

These gate-dependent features of the junction's critical current are accompanied by strong variations in the phase dependence, or shape, of the CPR (visible in Figure \ref{fig:two}a).  The most prominent feature of these shape variations is their forward skew, meaning that the critical current occurs at an advanced phase compared to a sine wave.  The skew of a CPR can be characterized by taking its Fourier transform and defining a quantity $S = -A_2/A_1$, where $A_1$ and $A_2$ are the first and second harmonics of the CPR.  Positive $S$ indicates forward skew, and hereafter we use $S$ and `skew' interchangeably.

Plotting the skew on the same gate voltage axis as the critical current reveals an intricate interplay between the shape of the CPR and its amplitude (light blue in Figure \ref{fig:two}b).  At the most negative gate voltages, recurring peaks in the critical current are accompanied by coincident peaks in the skew.  In the intermediate regime of fluctuating critical current, the skew also fluctuates and some correlations are observable between critical current and skew.  As the critical current begins to increase sharply at the most positive gate voltages, the skew remains relatively constant near $S=0.3$.  Throughout the gate trace, it is remarkable that $S$ almost never approaches $0$. The CPR is nearly always forward-skewed, in sharp contrast with the sinusoidal behavior ($S = 0$) found in the conventional Josephson effect.  At certain gate voltages, the skew is not calculated due to the CPR amplitude being too small to obtain a reliable result.

\section*{Short junction theory and CPR skew reveal mode number and transparency}

The variations of the CPR amplitude and shape with gate voltage contain considerable information about the electronic structure of the junction.  The CPR depends on the electronic properties of the junction according to the fundamental relation

\begin{equation}
   I(\phi) = \frac{2e}{\hbar}\frac{dF}{d\phi},
    \label{eq:freenergy}
\end{equation}

\noindent where $F$ is the free energy and $\phi$ is the phase difference across the Josephson junction \cite{beenakker1992sqpc}. In the limit where the junction length is short compared to the superconducting coherence length, the dominant contribution to the supercurrent comes from discrete spectrum of bound states in the junction, with energies $\epsilon_p = \Delta\sqrt{1-\tau_p\sin^2(\phi/2)}$ \cite{beenakker1991universal}. The energies $\epsilon_p$ of these states, called Andreev bound states, are related to the transmission probabilities $\tau_p$ of the underlying normal conduction channels in the junction. Here $\Delta$ is the proximity-induced gap. With this spectrum, Eq. (\ref{eq:freenergy}) can be simplified to a well-known formula \cite{kulik1975contribution,haberkorn1978cpr,zaitsev1984quasiclassical}:
\begin{align}
   I(\phi) &= -\frac{2e}{\hbar}\sum_{p=1}^N \frac{d\epsilon_p}{d\phi} \tanh\left(\frac{\epsilon_p}{2k_BT}\right)=\frac{e\Delta^2}{2\hbar}\sin{\phi}\sum_{p=1}^N\frac{\tau_p}{\epsilon_p}\tanh\left(\frac{\epsilon_p}{2k_BT}\right) . \label{eq:standardcpr}
\end{align}
Here $N$ corresponds to the number of Andreev bound states in the junction.

This simplified model illustrates that the CPR is a sensitive probe of both the number of modes and their transparencies.  In the simplest case of a single channel, the amplitude and skew of the CPR both increase with $\tau_1$.  In the limit where $\tau_1 \ll 1$, a sinusoidal CPR is expected ($S \to 0$), while the maximum skew $S=0.4$ occurs for $T=0$ K and $\tau_1=1$.  When more than one channel contributes to the CPR, the shape and amplitude of the CPR become sensitive to a set of $N>1$ transmission probabilities.

We performed fits of Eq. (\ref{eq:standardcpr}) to the CPR data, allowing up to $N=6$ contributing modes (See Supplemental Material \ref{sup:fit} for the fitting procedure).  Below $V_C \approx -1.16$ V, the CPR is well-described by a single contributing channel (dashed red line in Figure \ref{fig:two}b). An example of a single-mode CPR with high skew, measured at $V_C=-1.22$ V, is shown in black in Figure \ref{fig:two}c. The best fit of Eq. (\ref{eq:standardcpr}) to the data, in red, is consistent with a single mode with $\tau_1=1$ and $\tau_p = 0$ for $p>1$. 

In addition to fits, the single-channel regime is identified by a fixed relationship between the CPR amplitude and skew.  As shown in blue in Figure \ref{fig:two}d, calculated for $T=0$ and $\Delta = 128$ $\mu$eV, increasing critical current corresponds to increased skew, up to the maximum skew $S=0.4$.  A scatter plot (red dots in Figure \ref{fig:two}d) of data taken with the cutter gate voltage $V_C<-1.16$ V shows that the critical current and skew are highly correlated as one would expect for a single channel.  Experimentally this regime contains both the least and most skewed CPRs that we observed, with values as high as $S\approx 0.39$. The strong correlation between critical current and skew provides additional evidence for single-channel behavior.

Throughout the regime of fluctuating critical current, $V_C\approx-1.16$ V to $V_C\approx -0.45$ V, more channels contribute with fluctuating transmission probabilities. In Figure  \ref{fig:two}e, the measured CPR at $V_C=-0.9$ V is plotted in black along with a best fit in red to Eq. (\ref{eq:standardcpr}) with $N=2$ modes.  It is notable that the CPRs in Figure \ref{fig:two}c and \ref{fig:two}e have very similar amplitudes, but markedly different shapes, directly reflecting the contribution of multiple modes in Figure \ref{fig:two}e.  Hence, at $V_C=-0.9$ V the best fit requires a minimum of two modes with transparencies $\tau_1=0.95$ and $\tau_2=0.14$, whereas a single-mode fit is inconsistent with the experiment.

When more than one channel contributes to the CPR, there is no correlation between CPR amplitude and skew, in contrast with the single-channel regime.  Instead, a range of critical currents are possible for a given value of the skew, with the single-channel behavior providing a lower bound.  In Figure \ref{fig:two}f, a scatter plot of skew and critical current data over the entire cutter gate range confirms this expectation: data where $V_C > -1.16$ V (open black dots) occurs at larger critical currents than data where $V_C < - 1.16$ V. The lack of strong correlation between critical current and skew when $V_C > -1.16$ V is evidence that this regime is characterized by multiple channels.  The observed scatter captures the overall device behavior and constrains our numerical simulations, as we discuss below.

Above $V_C\approx-0.45$ V, additional channels enter with transmission probabilities which increase rapidly with gate voltage.  Interestingly, the CPR deviates from the short junction model at two of the peaks below $V_C = -1.16$ V, which we discuss further below in the context of electron-electron interactions.

\section*{Using numerical simulations to study the interplay of electrostatic potential, spatial structure, and spectrum of Andreev states}

\begin{figure}
   \centering
   \includegraphics[width=\textwidth]{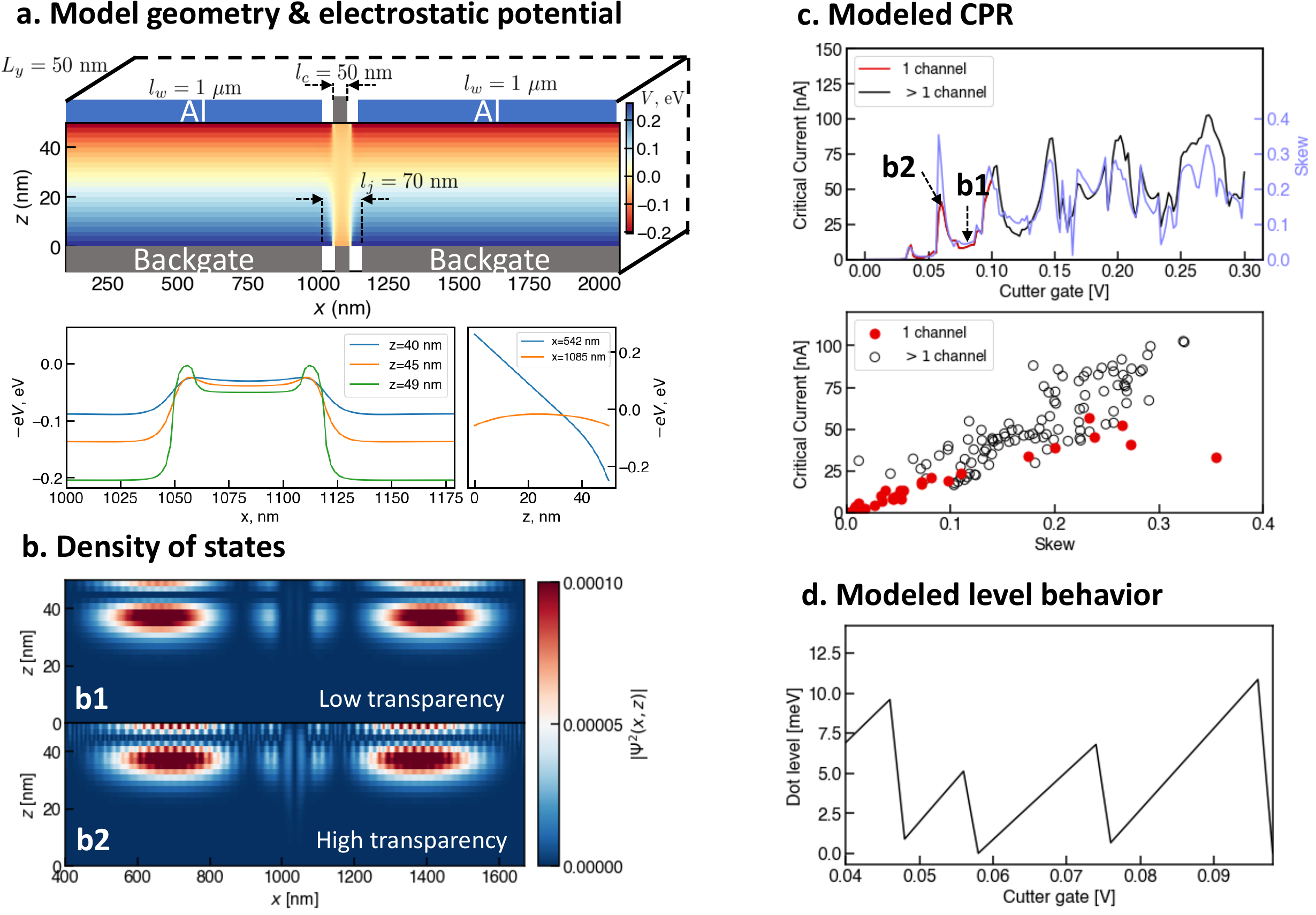}
   \caption{Numerical simulations of the InAs junction. a) Top: schematics of the simulated model, consisting of two wire segments and the junction, each augmented with an electric gate. Color indicates the calculated electrostatic potential for the left/right gate voltage $-0.29$~V and cutter voltage $0.06$~V. Bottom panel: 2d cuts of the electrostatic potential along the x direction (left) and along the z direction (right). b) The density of the lowest positive Andreev state at phase $\phi=\pi$ in the junction at cutter voltage $0.07$V (top, b1) and $0.06$V (bottom, b2), corresponding to the regimes of low and high transparency in the junction respectively. c) Top: Calculated dependence of the critical current (red/black) and skew (blue) on the cutter gate voltage. Arrows indicate the gate voltages of b1, b2 plots. Bottom: The dependence of the calculated critical current on the skew. In both panels red color indicates the single-channel regime. d) The dependence of the position of the highest occupied level in the junction on the cutter gate voltage. }
   \label{fig:three}
\end{figure}

Although the short junction model provides an intuitive description of the number and transparency of modes in the junction, it is limited in its ability to faithfully represent device behavior.  In a real device, the spectrum and spatial distribution of states will be influenced by the device geometry, material composition, and details of the electrostatic potential \cite{antipov2018gateeffects,woods2018gateeffects,Winkler2018}.  Understanding these effects is important for device engineering in qubit and Majorana applications, motivating a more detailed model of the observed CPR.

We study a microscopic model of the system, consisting of two wires and a junction, the geometry shown in the Figure \ref{fig:three}a. The wire's state is controlled by three electric gates: two below the wires and the cutter gate in the junction. In the normal state it is described by the following low-energy Hamiltonian
\begin{equation}\label{eq:H_n}
\hat H_n = \frac{1}{2m^*} \left(\hat k_x^2 + \hat k_y^2 + \hat k_z^2\right) \\
 \!+ V(x,z)\! -\! \alpha \left(\hat k_x \sigma_y \!-\! \hat k_y \sigma_x\right),
\end{equation}
where operators $\hat k_{x,y,z}$ denote the momentum in $x,y,z$ directions,  $\sigma_{x,y,z}$ are Pauli matrices, $m^*=0.026$~$m_e$ \cite{lutchyn2018review} is the effective mass in InAs ($m_e$ being the electron mass), $\alpha = 0.01$~eV$\cdot$nm is the spin-orbit coupling \cite{lutchyn2018review} and $V$ is the electrostatic potential, created by the electric gates. The potential is translationally invariant along the $y$ direction and can be calculated solving the Poisson equation $\nabla^2 V(x,z) = -\frac{e n(x,z)}{\varepsilon_0\varepsilon_r}$ with $\varepsilon_r = 15.2$ and the boundary conditions set by the voltages of the surrounding gates and the band-offset at the interface of InAs and Al, which we take as $W=0.25$~eV. In order to obtain the potential one has to self-consistently determine density of electrons and the potential in a Schrodinger-Poisson loop, which is computationally prohibitive. Instead we employ the Thomas-Fermi approximation for the density
$n = \frac{1}{3\pi^2}  \left(\frac{2 m^* \phi}{\hbar^2}\right)^{\frac{3}{2}},
$
which produces excellent agreement for the potential with the full Schrodinger-Poisson method \cite{antipov2018gateeffects,Mikkelsen2018hybridization}.

Superconductivity is included using a Bogoliubov-de-Gennes Hamiltonian and adding pairing potentials $\Delta$ and $\Delta e^{i\phi}$ to the left and right proximitized segments of the wires respectively, leading to the phase difference $\phi$ across the junction. The numerical complexity is alleviated by projecting the Hamiltonian in the $y$ direction to the basis of sinusoidal eigenstates \cite{antipov2018gateeffects,Stanescu2011majorana}. 

The zero-temperature CPR is calculated from the Andreev spectrum by taking the derivative of the ground state energy with respect to $\phi$ (see Eq. (\ref{eq:freenergy})) \cite{Meng2012}. The gate dependence of the simulated critical current is shown in the upper panel of Figure \ref{fig:three}c) (black), and is qualitatively similar to the experimental data. As the gate voltage increases, an insulating regime transitions into three well-separated peaks, which then evolve into a regime of fluctuating critical current. Within the regime of recurring peaks, the simulated supercurrent is dominated by the contribution from a single mode.
The skew of the modeled CPR tends to increase and decrease with the critical current (light blue in Figure \ref{fig:three}c). Plotting the simulated skews and critical currents against one another reveals that this relationship strongly resembles the experiment (lower panel in Figure \ref{fig:three}c). The simulated behavior with only one contributing mode, plotted in red, shows strong correlation between critical current and skew. These single-channel points lie predominantly along the bottom part of the scatter plot, as found experimentally.

The excellent agreement between the measured and simulated CPRs provides an opportunity to further examine the electronic structure of the device. In Figure \ref{fig:three}b, the local density of states in the single-channel regime reveals a striking difference in the structure of low and high transparency states. In both cases the band offset between aluminum and InAs leads to an enhanced density of states in the nanowire near the aluminum coating. At low transparency, the density of states in the junction area is low relative to that of the leads. In contrast, at high transparency the density of states becomes much larger in the junction. The change between regimes occurs in the range of $10$~mV difference in $V_C$, indicating the delicate nature of the quantum states in the junctions formed by InAs nanowires.

Figure \ref{fig:three}d shows the evolution with cutter gate voltage of highest occupied level energy in the junction, where zero energy corresponds to the chemical potential of the leads. This simulation demonstrates that sweeping the cutter gate voltage can be expected to bring a series of states in and out of resonance with the leads. In the single-channel regime in particular, this resonance behavior contributes to the formation of recurring peaks in the critical current.

\section*{Interacting behavior in the single-mode peaked regime}

\begin{figure}
   \centering
   \includegraphics[width=\textwidth]{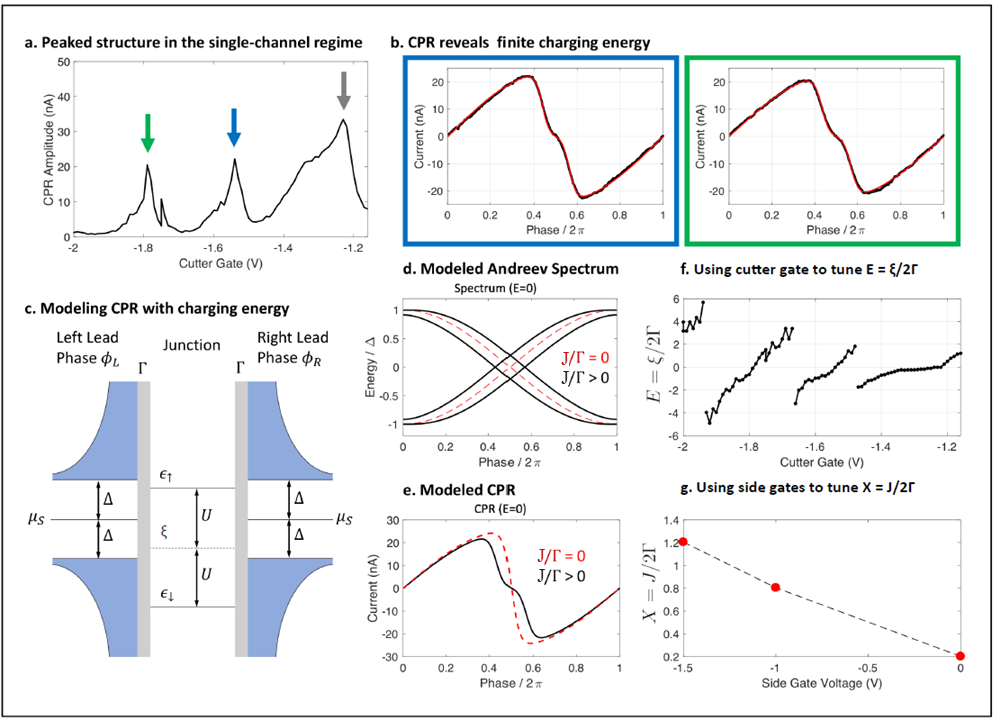}
   \caption{Using the CPR to elucidate the role of interactions in the single-channel regime.  a) In the single-channel regime, the CPR amplitude develops recurring peaks, indicated by arrows.  b) At two of these peaks, a shoulder appears in the CPR near phases equal to $\pi$ modulo $2\pi$.  The colored borders surrounding the plots correspond to colored arrows in (a).  The appearance of the shoulders in the CPR is evidence for finite charging energy in the junction region of the device.  c) Modeling our device with finite charging energy allows calculation of CPRs in this regime.  The two leads, with gap $\Delta$ and chemical potential $\mu_s$, are coupled to the junction symmetrically with coupling $\Gamma$.  The charging energy is modeled as an exchange energy $J$ in the device, which splits a state at energy $\xi$ in the junction into an odd and even parity state with respective energies $\epsilon_\downarrow$ and $\epsilon_\uparrow$.  d, e) The model allows calculation of the spectrum of Andreev states and the CPR, shown here without charging energy (red dashed), and with finite charging energy (black).  The appearance of a shoulder is characteristic of finite charging energy when $\xi/\Gamma \approx 0$, and can be used in fits to the data (red curves in (b)).  f) Fitting the model to the data throughout the single-channel regime reveals the evolution of energy levels in the device.  g) The model can also be fit to the data at different values of side gate voltage, revealing that more negative side gate voltage corresponds to increased $J/\Gamma$.}
   \label{fig:four}
\end{figure}

We now focus on the single-channel regime, to investigate in more detail the series of recurring peaks in critical current (Figure \ref{fig:four}a).  Near two of these peaks, the CPR deviates from the non-interacting behavior and instead displays a shoulder near phase $\phi=\pi$ (Figure \ref{fig:four}b).  When the cutter gate voltage $V_C$ is tuned away from the peaks, the shoulder feature is no longer observed (not shown, refer to Figure \ref{fig:two}), and the CPR reverts back to a shape consistent with non-interacting short junction theory.  The shoulder features are a precursor to $\pi$-junction behavior resulting from Coulomb blockade effect\cite{spivak1991cpr, rozhkov1999cpr, vecino2003cpr, siano2004cpr, choi2004cpr, sellier2005cpr, karrasch2008cpr, meng2009cpr, kirsanskas2015prb}, and have been studied experimentally in a carbon nanotube Josephson junction \cite{delagrange2015cntcpr}. Related phenomena have also been observed in nanowire-superconductor hybrids \cite{chang2013spec, lee2013spec, vandam2006screversal, szombati2016phi, spanton2017cpr}, but without quantitative modeling of the CPR.

To understand the shoulder behavior in the CPR, we consider a scenario where a quantum dot (QD) is formed in the junction. The QD is characterized by the strength of the local Coulomb electron-electron interaction $U$ and a single spin-degenerate level with energy $\xi$ measured relative to the chemical potential of the leads (see Figure \ref{fig:four}c). Coupling strengths to the right/left leads $\Gamma_{R/L}$ and the energy $\xi$ depend on the cutter gate voltage. Following Ref.~\cite{vecino2003cpr} we treat interactions within the mean-field approximation and introduce a local exchange field $J=U(\langle n_{\downarrow} \rangle - \langle n_{\uparrow} \rangle)/2$ where $\langle n_{\uparrow/\downarrow} \rangle$ are single-level occupation probabilities for spin-up and spin-down electrons. In the limit when $\Gamma_R=\Gamma_L=\Gamma \gg \Delta$, the resulting Andreev energies can be obtained analytically:\\

\begin{equation}
   \epsilon_\pm = \Delta \sqrt{\frac{\cos^2\phi/2+2E^2+Z^2(Z^2+\sin^2\phi/2) \pm 2XS(\phi)}{Z^4+2(X^2+E^2)+1}}.
   \label{eq:dotspectrum}
\end{equation}

\noindent Here $\phi$ is the phase difference $\phi_L-\phi_R$ between the left and right leads, 
$S(\phi) = \sqrt{Z^2\cos^2\phi/2+E^2+(\sin^2\phi)/4}$, $E=\xi/2\Gamma$, $X=J/2\Gamma$, and $Z^2=X^2-E^2$.  With the above formula for the Andreev spectrum, we have the following expression for the current-phase relation:

\begin{equation}
   I(\phi)=-\frac{e}{\hbar}\left(\tanh\left(\frac{\epsilon_+}{2k_BT}\right)\frac{d\epsilon_+}{d\phi} + \tanh\left(\frac{\epsilon_-}{2k_BT}\right)\frac{d\epsilon_-}{d\phi} \right).
   \label{eq:dotcpr}
\end{equation}

In this model, the expected junction behavior results from an interplay between the level energy $\xi$ and the on-site exchange coupling $J$. The characteristics peaks in the supercurrent  observed in the experiment, see Figure \ref{fig:four}a, are related to the charge-degeneracy points. We now analyze CPR using a simple model above and consider $\xi/\Gamma=0$ case. Without interactions, such a situation corresponds to a resonant transmission through the QD with CPR given by the red line in Figure \ref{fig:four}d. The exchange interactions effectively split spin-degenerate level creating level crossings away from $\phi=\pi$ (black curves in Figure \ref{fig:four}d). Using this spectrum one finds the appearance of a shoulder near $\phi=\pi$ in CPR. By fitting the experimental data to the Eq. \eqref{eq:dotcpr}, we find a good agreement with the measured CPR data, with best fit values for the interaction strength given by $J/\Gamma=0.39$ and $0.42$ when $V_C=-1.54$ V and $-1.79$ V respectively (red dashed curve in Figure \ref{fig:four}b is the best fit).            

Away from the single-particle resonance (i.e. $\xi \neq 0$), the dispersion of the Andreev states with $\phi$ gets suppressed due to the reduction in transmission probability through the QD.  A nonzero exchange energy splits the Andreev level spin-degeneracy as before, but as long as $\xi \gg J$ the levels will not cross at zero energy. Thus, one expects the reversal of the supercurrent to appear in the CPR only when the junction is tuned near resonance. Indeed, we only observe shoulders in the CPR near peaks in the critical current, consistent with the theoretical predictions. 

Fitting the measured CPR to Eq. (\ref{eq:dotcpr}) throughout the single-channel regime allows one to extract the cutter gate dependence of the normalized site energy, $\xi/\Gamma$ (Figure \ref{fig:four}f).  As the cutter gate increases from $-2$ V to $-1.16$ V, three levels are tuned in and out of resonance with the leads.  In addition to shifting the spectrum, the cutter gate is expected to modify the shape of the confining potential in the junction and the coupling $\Gamma$.  We speculate that the coupling $\Gamma$ may increase with $V_C$, contributing to the decrease in slopes in Figure \ref{fig:four}f as $V_C$ becomes more positive.  The experimentally extracted evolution of levels agrees qualitatively with numerical simulations (see Figure \ref{fig:three}d).

In the measurements discussed so far, only the cutter gate voltage was varied, with all other gates grounded.  With negative voltage $V_S$ applied to the two side gates, we observed additional CPRs with shoulder-like features.  Fitting these CPRs to Eq. (\ref{eq:dotcpr}) yielded values for the normalized exchange energy $J/\Gamma$.  As the side gate voltages become more negative, we find an increase in the mean value of $J/\Gamma$ at each gate voltage (Figure \ref{fig:four}g and Supplemental Material \ref{sup:gates}).  Fringe electric fields at the two ends of the junction may lead to decreased coupling $\Gamma$ with more negative $V_S$, consistent with this finding.  We observed no shoulder-like features in any CPR with positive side gate voltage.

\section*{Conclusion}

Our measurements and analysis distinguish between a multi-channel regime and a single-channel regime with widely tunable anharmonicity and interactions.  Our CPR-based technique for comparing experiment and numerical simulations allows detailed study of the interplay between mode structure and electrostatics, and may be extended to the design and analysis of future superconductor-semiconductor heterostructures \cite{vaitiekenas2016sag}. The propensity for quantum dot formation in a nanowire-based junction may also be viewed as a resource to detect Majorana modes. For example, a recent analysis found that $\pi$-junction behavior occurs with a quantum dot between two conventional superconducting leads~\cite{vandam2006screversal}.  

The presence of interactions at low electron density also has implications for experiments examining mode behavior in superconductor-nanowire hybrids, where interactions are sometimes neglected.  For example, a recent work examined multiple Andreev reflections as a probe of transparency in few-mode junctions, but ignored interaction effects which are known to influence IV characteristics \cite{goffman2017mar, avishai2001dot}.  Spectroscopy experiments also studied anharmonicity and Zeeman-induced spin-splitting of Andreev levels in the few-mode regime, but may also be affected by splitting effects from interactions as examined in our work \cite{vanwoerkom2017uwave, kringhoj2018anharmonicity}.  Overall our ability to accurately model device behavior, as well as to understand and control both interactions and anharmonicity, will elucidate the interpretation of current experiments and provide a method for device design.

\bibliographystyle{naturemag}
\bibliography{cprbib}

\section*{Acknowledgement}
We thank S. Vaitiek\ifmmode \dot{e}\else \.{e}\fi{}nas, J. Kirtley and G.~Winkler for useful discussions. The scanning SQUID measurements were supported by the Department of Energy, Office of Basic Energy Sciences, Division of Materials Sciences and Engineering, under Contract No. DE-AC02-76SF00515. Nanowire growth and device fabrication was supported by Microsoft Project Q, the Danish National Research Foundation, the Lundbeck Foundation, the Carlsberg Foundation, and the European Research Commission through starting grant HEMs-DAM, grant no.716655. S. Hart acknowledges support from the GLAM Postdoctoral Fellowship at Stanford University. C.M. Marcus acknowledges support from the Villum Foundation.

\newpage

\renewcommand{\thefigure}{S\arabic{figure}}
\setcounter{figure}{0}

\section*{Supplemental Material}

\subsection{Estimating the Amplitude of the Critical Current}\label{sup:mutual}
The advantage of having a flux-coupled measurement (being contact-less) of the Josephson current-phase relation also comes with the drawback of lacking a direct measurement of the amplitude of the current. In order to convert the measured flux signal into a phase-dependent Josephson current, we model the SQUID-sample geometry to the best of our knowledge to obtain the mutual inductance between the sample ring and the SQUID pickup loop.

First we apply a current $I_{FC}$ into the SQUID field coil to introduce a phase drop of $\phi = 2\pi \cdot \frac{I_{FC}\cdot M_{R-FC}}{\Phi_0}$ across the Josephson junction. $M_{R-FC}$ is the mutual inductance between the sample ring and the SQUID field coil, set by the measurement geometry between the SQUID and the sample. $\Phi_0$ is the superconducting flux quantum. We estimate the kinetic inductance of the Josephson junction to be much smaller than the self-inductance of the sample ring. Therefore, the persistent current due to fluxoid quantization is small, and thus the preceding equation is valid. Since the measured signal must be $2\pi$-periodic and $I_{FC}$ is known, we can easily extract $M_{R-FC}$ from the data.

The phase drop across the Josephson junction leads to a supercurrent $I(\phi)$ that we inductively measure. The measured signal in the pickup loop $\Phi_{PU}$ is related to the actual supercurrent by $\Phi_{PU}$ = $I(\phi) \cdot M_{R-PU}$. To estimate $M_{R-PU}$, we model the field coil, pickup loop and sample ring as three coaxial 1-dimensional circles. The distance between the field coil and pickup loop is fixed to be 280 nm from the fabrication process \cite{kirtley2016scanning}. The distance from the sample to the pickup loop is unknown and estimated to be 1.6 $\mu $m. Using Maxwell's analytic formula for two coaxial circles we can calculate the pair-wire mutual inductance \cite{NISTmutual}:
\begin{align}\label{eq:maxwell}
    M(r_1,r_2,h) &= \mu_0 \sqrt{r_1 \cdot r_2} \left[ \left(\frac{2}{k} - k \right) \cdot K - \frac{2}{k} \cdot E \right] \\
    k &= \frac{2 \cdot \sqrt{r_1 \cdot r_2}}{\sqrt{\left( r_1 + r_2 \right)^2 + h^2}} \nonumber
\end{align}

Where $\mu_0$ is the vacuum permeability, $r_1$ and $r_2$ are the radii of the two circles, $h$ is the distance between their centers, K and E are respectively the complete elliptic integrals of the first and second kind with modulus k. 

\begin{figure}
   \centering
   \includegraphics[width=\textwidth]{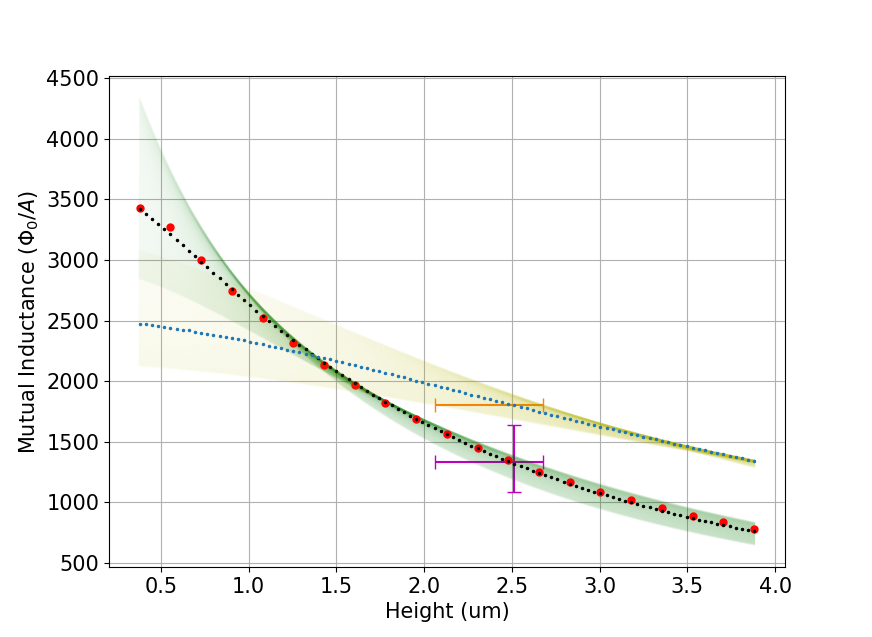}
   \caption{Calculated mutual inductance between the field coil and the sample ring ($M_{R-FC}$) and between the sample ring and the pickup loop ($M_{R-PU}$), as a function of height from the pickup loop to the sample ring. \textbf{Blue dots}: calculated $M_{R-FC}$ with Eq. (\ref{eq:maxwell}), field coil effective radius = 7.07 $\mu$m and sample ring effective radius = 4.03 $\mu$m. \textbf{Yellow shade}: range of $M_{R-FC}$ with field coil radius between 6 and 8 $\mu$m. \textbf{Black dots}: calculated $M_{R-PU}$ with Eq. (\ref{eq:maxwell}), sample ring effective radius = 4.03 $\mu$m and pickup loop effective radius = 3.26 $\mu$m. \textbf{Red dots}: superconducting thin film simulation of $M_{R-PU}$ with method in \cite{Brandt2005}. \textbf{Green shade}: range of $M_{R-PU}$ with sample ring radius between 3.5 and 4.5 $\mu$m. \textbf{Orange error bar}: height uncertainty at $M_{R-FC}$ = 1800 $\frac{\Phi_0}{A}$. \textbf{Purple error bars}: projected height uncertainty results in an estimated range of $M_{R-PU} \sim$ 1100 $-$ 1600 $\frac{\Phi_0}{A}$.}
   \label{fig:sup_mutual}
\end{figure}

To compute the pair-wise mutual inductance at various heights (black and blue dotted lines in Figure \ref{fig:sup_mutual}), we use the effective radii of each superconducting annulus 
defined by $r_{\rm eff} = \sqrt{\frac{r_{\rm inner}^2 + r_{\rm outer}^2}{2}}$. The calculated $M_{R-PU}$ is also found to be in very good agreement with a thin film superconductor model (red dots) \cite{Brandt2005}. Using an experimentally extracted $M_{R-FC}$ = 1800 $\frac{\Phi_0}{A}$, we find the real height to be $\sim$ 2.5 $\mu$m (0.9 $\mu$m off from the nominal height probably due to obsolete scanner calibration). To take into account the systematic uncertainty in the current distribution in the field coil and sample ring, we also compute the range of mutual inductance resulting from any arbitrary current distribution in the field coil and the sample ring (yellow and green bands respectively in Figure \ref{fig:sup_mutual}). Subsequently we obtain an uncertainty in height at $M_{R-FC}$ = 1800 $\frac{\Phi_0}{A}$ (orange error bar in Figure \ref{fig:sup_mutual}). We project the height uncertainty onto $M_{R-PU}$ and obtain a range $\sim$ 1100 $-$ 1600 $\frac{\Phi_0}{A}$. Additionally, off-axial alignment can give $\pm$10$\%$ of uncertainty in $M_{R-PU}$. We have processed the data in this paper with $M_{R-PU}$ = 1170 $\frac{\Phi_0}{A}$, within the range of our systematic uncertainty and consistent with previous scanning SQUID measurements \cite{spanton2017cpr,sochnikov2013direct,sochnikov2015nonsinusoidal}.

\subsection{Effects of Different Electrostatic Gates}\label{sup:gates}
\begin{figure}
   \centering
   \includegraphics[width=0.8\textwidth]{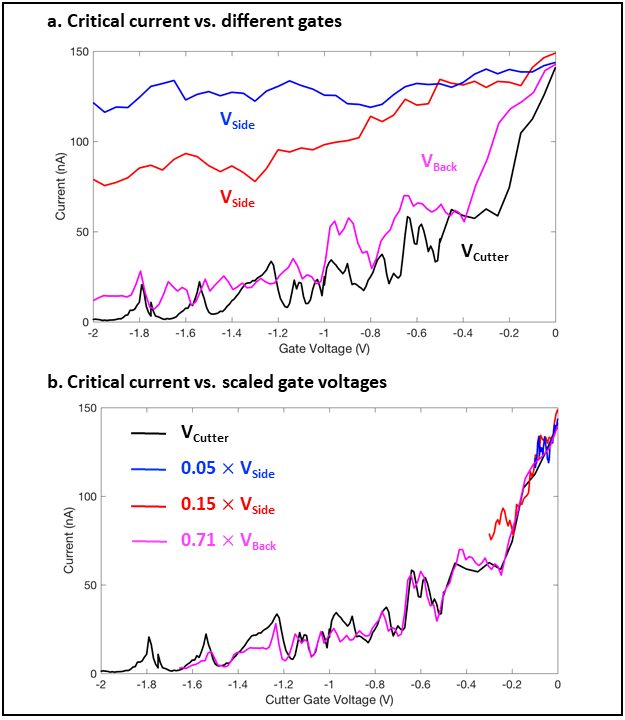}
   \caption{Evolution of the Josephson critical current (measured as the CPR amplitude) with different gate voltages. a) Each of the four gates is varied independently while keeping all three other gates and the aluminum ring grounded. The back gate and cutter gate both deplete the semiconductor junction at negative gate voltages, whereas the side gates only show a mild effect on the junction. b) The side gate and back gate voltages are scaled by arbitrary factors to best match the colored curves to the black (cutter gate) curve. The purple and black curves (back gate and cutter gate respectively) match very well, indicating that the back gate acts simply as a less effective cutter gate due to electrostatic screening from the grounded epitaxial aluminum shell.}
   \label{fig:sup_gates}
\end{figure}

\begin{figure}
   \centering
   \includegraphics[width=\textwidth]{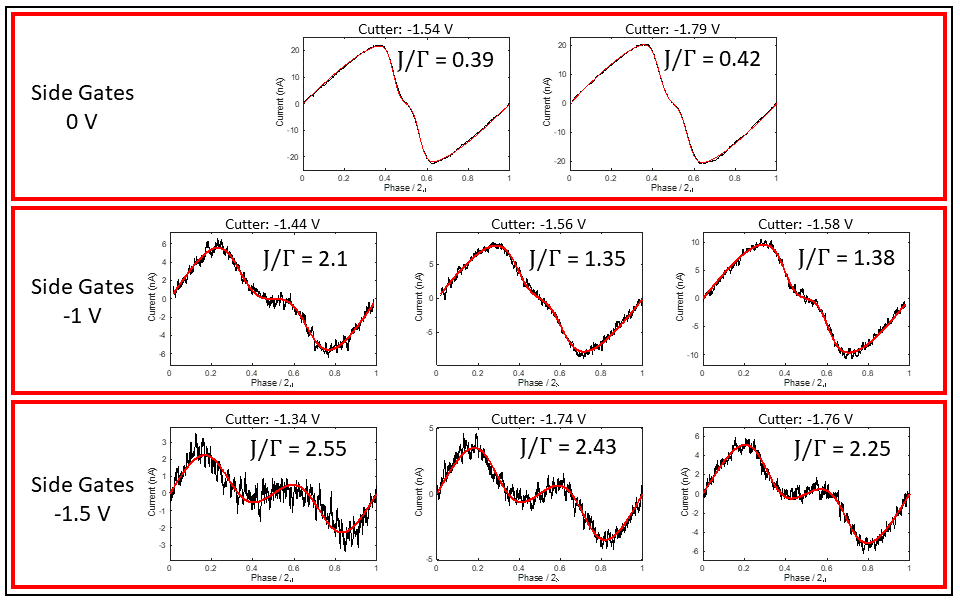}
   \caption{Normalized exchange energy $J/\Gamma$ in the interacting regime as a function of side gates and cutter gate voltages, as extracted from fitting. Black lines are experimental data. Red lines are fits to Eq. \ref{eq:dotcpr} with four free parameters: the CPR amplitude (to account for systematic uncertainties, see Supplemental Material \ref{sup:mutual}), temperature,  $E = \xi/2\Gamma$, and $X = J/2\Gamma$. The two side gates are shorted together. }
   \label{fig:sup_gamma_vs_sidegate}
\end{figure}

We measure current-phase relation as a function of each of the four electrostatic gates fabricated on the device (two side gates, one cutter gate and one back gate, as shown in Figure \ref{fig:one}). We find that all four gates have different effects on the evolution of the CPR amplitude (shown in Figure \ref{fig:sup_gates}a). When we scale the back gate voltages by an arbitrary factor of 0.71, we see that the effect of the back gate is qualitatively similar to that of the cutter gate (Figure \ref{fig:sup_gates}b). We speculate that the epitaxial aluminum shell screens the electric field everywhere along the wire except the junction region, resulting in the back gate having the same effect on the juntion behavior as the cutter gate. On the other hand, the two side gates only cover parts of the wire with epitaxial aluminum, thus are strongly screened and cannot fully deplete the junction region (see Figure \ref{fig:sup_gates}a and b). The observation of different gating behavior signifies the importance of understanding the spatial structure of the electrostatic potential and the bound states, an issue we address by numerical simulations of the device geometry in the main text. 
Even though the side gates do not effectively deplete electrons in the junction region, they affect the tunneling rates $\Gamma_{R/L}$ onto the junction in the interacting regime. Figure \ref{fig:sup_gamma_vs_sidegate} shows the effect of both the side gate voltage and the cutter gate voltage on the normalized exchange energy $J/\Gamma$. The electric field directly underneath the side gates are expected to be screened by the epitaxial aluminum, but the fringe electric fields creeping around the ends of the junction may lead to the systematic suppression of the tunneling rate $\Gamma$ at negative side gate voltages as seen in the experiment. Future device design may deliberately place finger gates over the two ends of the Josephson junction to further explore the interacting behavior.

\subsection{Model Fitting in the Single- and Multi-Channel Regime}\label{sup:fit}
We use the ``fit()'' function in MATLAB to perform a least-squares fit of the experimental data to the short-junction current-phase relation formula (\ref{eq:standardcpr}) in order to extract the number of independent channels and the corresponding transmission coefficients. 

We first use a one-channel model with 3 free parameters $\{\Delta, T, \tau\}$, respectively the superconducting gap in the leads, the temperature, and the transmission coefficient as defined in Eq. (\ref{eq:standardcpr}), to fit the most skewed CPR at V$_C$ = $-$1.22 V (see main text Figure \ref{fig:two}c). The high quality single-mode fit yields $\tau$ = 1, $\Delta$ = 128 $\mu$eV and T = 98 mK. We then use a two-channel fit with 4 free parameters $\{\Delta, T, \tau_1, \tau_2\}$ to fit the same data. For the two-channel fit, the $\chi^2$ 95$\%$ confidence interval constricts $\tau_2 \ll$ 0.01. Therefore we conclude that only one channel is contributing to the supercurrent at V$_C$ = $-$1.22 V. Fitting this single perfectly transmitting channel yields very constrained values of $\Delta$ and T, with 95$\%$ confidence intervals of  $\sim$ 2 $\mu$eV and $\sim$ 2 mK respectively, hence we fix $\Delta$ = 128 $\mu$eV and T = 98 mK for subsequent multi-channel fits. Notice that $\Delta$ = 128 $\mu$eV is smaller than the expected gap from 7 nm of epitaxial aluminum film \cite{deng2016boundstate,vaitiek2018gfactor}. This can be explained by the inability of the back gate to deplete the wires of electrons, which would make it easier to proximitize to the superconductor \cite{antipov2018gateeffects}, as well as large systematic amplitude uncertainty as shown in Supplemental Material \ref{sup:mutual}, in addition to effects from the evaporated (disordered) bulk aluminum ring.

\begin{figure}
   \centering
   \includegraphics[width=0.8\textwidth]{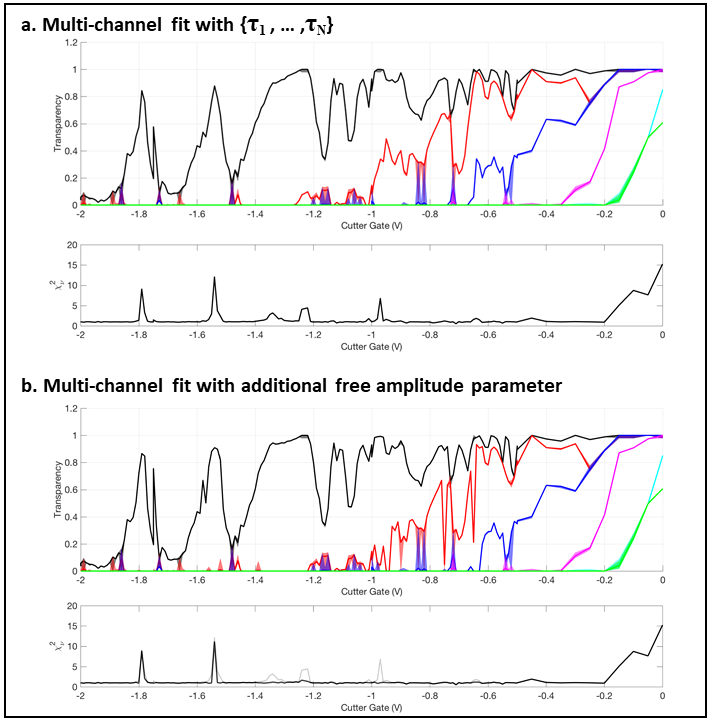}
   \caption{Fitted transmission coefficients and reduced $\chi^2$ statistics as a function of cutter gate voltage. a) Best-fit transmission coefficients and calculated $\chi^2_\nu$ as a measure of fit quality. Only transmission coefficients are free parameters (with fixed $\Delta$ = 128 $\mu$eV, T = 98 mK). Different colors represent independent channels contributing to the Josephson supercurrent. At each cutter gate voltage, the number of colored lines indicate the number of channels N giving the best fit to the data. The vertical (shaded) width of the colored lines indicate 68$\%$ confidence interval of the fitted parameter value. b) Adding an additional free parameter for the supercurrent amplitude noticeably improves $\chi^2_\nu$. Note that since $\chi^2_\nu$ takes into account the number of degrees of freedom, this is not a trivial improvement due to more free parameters.}
   \label{fig:sup_modes_chi2}
\end{figure}

For N-channel fits, we manually set the number of channels in the fit through N = 1, 2, 3 ... and use N free parameters $\{\tau_1, \tau_2,...,\tau_N\}$ to perform the least-squares fit to Eq. (\ref{eq:standardcpr}) at each value of N. The condition for the best fit is met when additional channels do not improve the fit quality, as measured by the residuals. The fitted transmission coefficients $\{\tau\}_N$ over the entire cutter gate voltage range is shown in Figure \ref{fig:sup_modes_chi2}. 68$\%$ confidence intervals for the fit parameters are calculated from standard $\chi^2$ contour plots \cite{bevington2003data}. The confidence intervals (i.e. error bars) for each fitted transmission coefficient is shaded in Figure \ref{fig:sup_gates}a. Notice these error bars only represent the statistical uncertainties in the fits and not the systematic uncertainties in the experiment.

Here we find our data to be consistent with up to 6 independent channels according to the short-junction theory. Below V$_C$ = $-$1.16 V, the CPR is dominated by a single channel. As V$_C$ becomes more positive, more channels are introduced and their transparencies fluctuate until V$_C$ $\sim$ $-$0.4 V, above which both the transparencies and the overall CPR amplitude monotonically increase towards positive cutter gate voltages. 

To further validate our analysis, we calculate the reduced chi-squared ($\chi^2_\nu$) statistic as a test for goodness-of-fit \cite{bevington2003data}. The reduced chi-squared statistic is defined as the following:
\begin{equation}\label{eq:chi2}
    \chi^2_\nu = \frac{\chi^2}{\nu}, \hspace{1cm} \chi^2 = \frac{1}{\sigma^2}\sum_i\left( O_i - C_i \right)^2, 
\end{equation}
where $\chi^2$ is the chi-squared statistic, $\sigma^2$ is the variance of the experimental noise distribution, $O_i$ is the measured data. In this case it is the measured Josephson current at each phase bias. $C_i$ is the theoretical model value. In this case it is the theoretical Josephson current at each phase bias given by Eq. \ref{eq:standardcpr}.$\nu$ is the degree of freedom of the fit, given by $\nu = n - p$, where $n$ is the total number of data points and $p$ is the number of fit parameters.

If the data consists of a known theoretical model with free parameters plus Gaussian-distributed experimental noise, a set of best-fit parameters should result in $\chi^2_\nu \sim 1$, indicating that the data is well-described by the fitted model. Our measured SQUID noise floor, in units of the Josephson current from a ring, indeed follows a Gaussian distribution with variance of $\sim$ 1 (nA)$^2$, which we use as an estimator for $\sigma^2$ in Eq. \ref{eq:chi2}. Using the residuals from the fits, the SQUID noise floor and the number of fit parameters, we calculate $\chi^2_\nu$ at each cutter gate voltage. The vast majority of our fits result in $\chi^2_\nu \sim 1$, indicating that the data is highly consistent with the short-junction model with independent channels (Figure \ref{fig:sup_modes_chi2}a).

\subsection{Deviations of the Fit from the Short-Junction Model}\label{sup_delta_v_gate}
At a few values of gate voltage, we find that $\chi^2_\nu \neq \sim 1$, indicating that the data cannot be fully described by the non-interacting short-junction model. Near peak transparency ($\tau \sim 1$) in the single-channel regime, the charging energy modifies the shape of the CPR as shown in Figure \ref{fig:four}b and Figure \ref{fig:sup_gamma_vs_sidegate}, thus resulting in bad fits to the non-interacting model (Eq. \ref{eq:standardcpr}). For V$_C$ > $-$0.2 V, more channels (> 4) begin to participate in the Josephson current, potentially causing the independent-channel assumption in the model to breakdown, resulting also in large $\chi^2_\nu$. In addition, we notice that between V$_C$ = $-$1.36 V and V$_C$ = $-$0.95 V, the $\chi^2_\nu$ indicates poor fit quality that is not clearly related to charging energy or interacting channels. 

We revisit our multi-channel fit with an additional free amplitude parameter (a prefactor close to 1 that scales the amplitude of the supercurrent) and obtain improved fit quality as indicated by $\chi^2_\nu$ (Figure \ref{fig:sup_modes_chi2}b). We interpret the remaining large peaks in $\chi^2_\nu$ as due to either charging energy or multi-channel interactions. Although the free amplitude parameter, which varies up to $\pm$ 10\%, is equivalent to a changing $\Delta$,  We are reluctant to consider this a measurement of a gate-voltage dependent superconducting gap. Weak interactions and geometric effects can cause subtle deviations in the shape of the CPR from the short-junction theory. In particular, the real device consists of proximitized superconducting leads, forming an S-S'-N-S'-S junction (where S refers to the bulk aluminum, S' refers to InAs nanowire proximitized by the epitaxial aluminum shell, N refers to InAs nanowire without the epitaxial aluminum shell). In such a case, the spatial distribution of the eigenstates in the nanowire affects the superconductivity. Though absent in conventional SNS short-junction theories, this important aspect of the device is directly captured in our numerical modeling, which helps to illuminate the voltage dependence of mode behavior in our data.



\end{document}